\documentclass[]{article}

\usepackage{mathtools}

\DeclarePairedDelimiter{\evdel}{\langle}{\rangle}
\DeclarePairedDelimiter{\bra}{\langle}{\rvert}
\DeclarePairedDelimiter{\ket}{\lvert}{\rangle}
\newcommand{\mel}[3]{\bra{#1}#2\ket{#3}}
\newcommand{\comm}[2]{\lbrack#1,#2\rbrack}

\date{March 31, 2017}
\title{Quantum operators for the computation of exponential weighted integrals of expectation values}
\author{Simone Sturniolo}

\begin{document}

\maketitle

\begin{abstract}
An analytically derived 'integral operator' approach is introduced to estimate the expectation value of a quantum operator for an evolving state weighted with an exponential function. This allows to compute quantities useful in Nuclear Magnetic Resonance or Muon Spectroscopy experiment simulations with greater speed and precision than the standard numerical integration approach.
\end{abstract}

\section{Introduction}

A common technique used in muon spectroscopy is avoided level crossing resonance \cite{HEMING,Cox1987}. In this kind of experiment, a beam of polarised muons is injected into a sample under a strong external magnetic field, and their subsequent decay causes emission of positrons whose momentum is directly related to the expectation value of $S_z$. The final spectrum is determined by the integral of this quantity over time, weighted with the decaying density of muons, for a given external magnetic field $B$:

\begin{equation}\label{eqn:alc}
I_{ALC}(B) \propto \int_{0}^{\infty}{\evdel{S_z}(B, t)e^{-at}dt}
\end{equation}

with $a=\frac{1}{\tau_\mu}$ here is the natural decay time of the muons. To compute the quantity seen in equation \ref{eqn:alc} in a simulation the most straightforward method involves evolving the state of the system with the time-dependent Schr\"{o}dinger equation, tracing out the eigenvalue and numerically integrating. In this note a different approach to compute this kind of integral is proposed. This involves finding an `integral operator' such that its expectation value will match exactly the result of equation \ref{eqn:alc}. It should also be noted that many similar problems can be described in the same general form; for example, if $a=i\omega$, we get the Fourier transform at a given frequency $\omega$. This method could thus apply to exploring specific frequencies of the spectrum of an expectation value without having to compute its full time evolution.

\section{Derivation}

Let us consider the case of a quantum system evolving under a general Hamiltonian. In the Schr\"{o}dinger picture, the wavefunction of such a system would be time dependent and the operator whose expectation value we wish to average (in this case, $S_z$) would be time independent. However we will be looking for a Heisenberg picture operator $O$ such that

\begin{equation}\label{eqn:intop}
\int_0^t\mel{\psi}{e^{\frac{iHt'}{\hbar}}S_ze^{-at'}e^{-\frac{iHt'}{\hbar}}}{\psi}dt' = \mel{\psi}{e^{\frac{iHt}{\hbar}}O(t)e^{\frac{-iHt}{\hbar}}}{\psi}-\mel{\psi}{O(0)}{\psi}
\end{equation}

where we took care to make the time dependence of the wave function explicit. By recalling the expression for the derivative of an expectation value in the Heisenberg picture we can write

\begin{equation}\label{eqn:opdev}
S_ze^{-at} = \frac{i}{\hbar}\comm{H}{O}+\frac{\partial O}{\partial t}
\end{equation}

It should be noted that since we moved from expectation values to operators this is not a necessary condition, but it is sufficient to guarantee that equation \ref{eqn:intop} holds.\newline
For systems with a finite number of states we can describe the operators as matrices without loss of generality. For convenience, let us consider a Hilbert space in which the Hamiltonian is diagonal and has eigenvalues $\lambda_i$. We will also rewrite $S_z$ as a general operator $S$, since we don't know anything about its form in this basis, and element-wise we can impose:

\begin{equation}\label{eqn:diffeq}
	s_{ij} e^{-at} = \frac{i}{\hbar}o_{ij}(\lambda_i-\lambda_j)
	+\dot{o}_{ij}
\end{equation}

where lowercase letters refer to elements of the correspondingly named uppercase operators. This differential equation can be solved easily. First we consider a solution for the associated homogeneous equation:

\begin{equation}\label{eqn:sol_hom}
\frac{i}{\hbar}o_{ij}(\lambda_{i}-\lambda_{j})+\dot{o_{ij}}=0\implies o_{ij}=o_{ij}^{0}e^{-i\frac{\lambda_i-\lambda_j}{\hbar}t}
\end{equation}

where we notice the presence of a matrix of integration constants $O^0$. Then we find a particular solution:

\begin{equation}\label{eqn:sol_part}
s_{ij}e^{-at}=\frac{i}{\hbar}o_{ij}(\lambda_{i}-\lambda_{j})+\dot{o_{ij}}\implies o_{ij}=\frac{s_{ij}}{\frac{i}{\hbar}(\lambda_{i}-\lambda_{j})-a}e^{-at}
\end{equation}

The integration constants may give us some thought, but in fact do not matter. Since this is a Heisenberg operator, in order to account for the time evolution it will need to be multiplied by a factor of $e^{i\frac{\lambda_i-\lambda_j}{\hbar}t}$. But for equation \ref{eqn:sol_hom} that means the entire term becomes a constant, and since equation \ref{eqn:intop} shows that the final value of interest only depends on the difference between the operator at time $t$ and $0$, such constants always cancel each other. Therefore equation \ref{eqn:sol_part} gives in fact the full solution we need. Including the time dependence factor we can define a new operator $P$ and write

\begin{equation}\label{eqn:full_op}
\int_{0}^{t}\mel{\psi(t)}{S_z e^{-at'}}{\psi(t)}dt' = \mel{\psi(0)}{P(t)}{\psi(0)}
\end{equation}

where 

\begin{equation}\label{eqn:full_op_elements}
p_{ij}(t) = \frac{s_{ij}}{\frac{i}{\hbar}(\lambda_{i}-\lambda_{j})-a}\left(e^{-at+i\frac{\lambda_i-\lambda_j}{\hbar}t}-1\right)
\end{equation}

\section{Performance}

\begin{figure}
	\includegraphics[width=\linewidth]{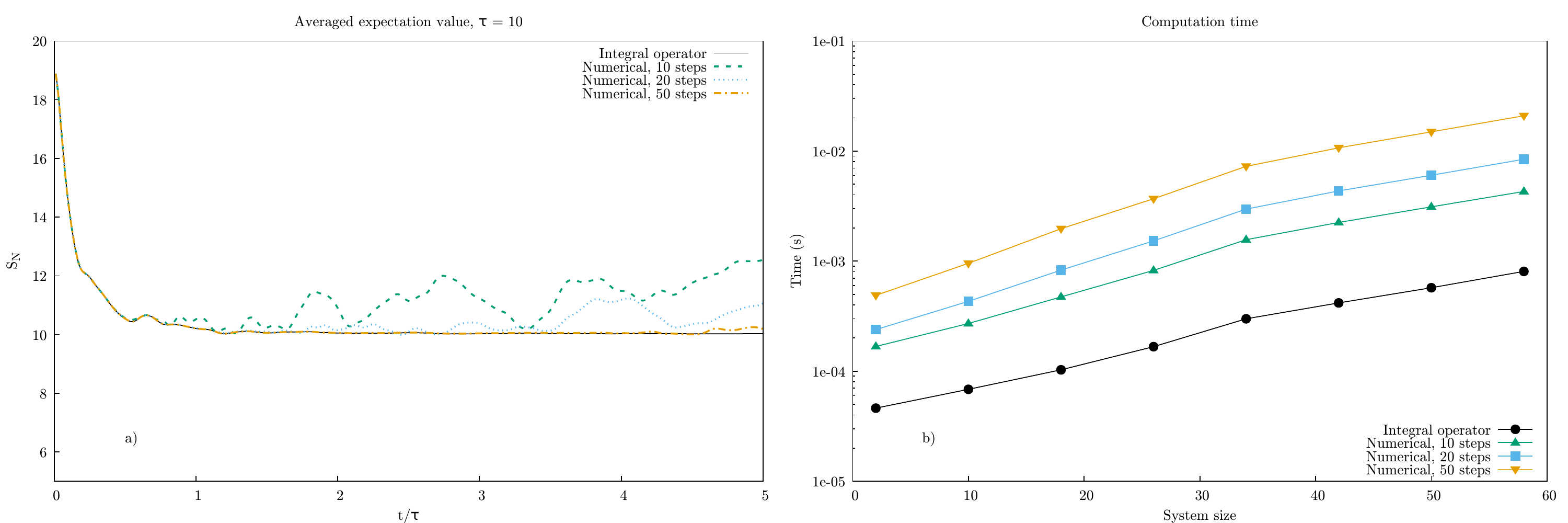}
	\caption{Time evolution for a random Hamiltonian of size 20 and computation times for random Hamiltonians with varying sizes and step numbers.}
	\label{fig:test1}
\end{figure}

We now proceed to test the performance of the proposed algorithm by generating random Hermitian Hamiltonians of variable size. The simulations were performed using Python and the numerical libraries Numpy and Scipy \cite{numpy, scipy} and run on a desktop PC. The script used to produce the data is given for reference as supplementary material and to provide an implementation of the algorithm. The generated Hamiltonians are all created as uniform random numbers between -0.5 and 0.5 for both real and imaginary part. Units are arbitrary. The first test focuses on evaluating the time evolution of the number operator $S_N$ for a density matrix initialized in the highest state and computing the weighted time average

\begin{equation}\label{eqn:timeavg}
\overline{S}_N = \frac{\int_{0}^{t}{\evdel{S_N}(t')e^{-\frac{t'}{\tau}}dt'}}
{\int_{0}^{t}{e^{-\frac{t'}{\tau}}dt'}}
\end{equation}

for $\tau=10.0$. Figure \ref{fig:test1}a shows the example of this integral computed at different times with the integral operator as well as with a numerical integration method for different amount of steps. The numerical integration was carried out with Numpy's default trapezoidal integration function; the number of steps is fixed for all points (meaning that the resolution is better at short times). The Hamiltonian diagonalization is not counted in the time as it is a necessary step for both methods. It is possible to compute the time evolution numerically without diagonalizing the Hamiltonian, but that introduces a further source of errors and requires smaller time steps, therefore is not desirable for small systems and has not been tried. It can be seen how increasing the number of steps makes the numerical solution progressively closer to the analytical result. In figure \ref{fig:test1}b we can however take a look at the average performance of these approaches, computed over 200 averages for different system sizes and step numbers. It becomes obvious that even for the lowest integration step counts the numerical approach is far inferior to the integral operator, taking almost an order of magnitude longer for worse results.

\begin{figure}
	\centering
	\includegraphics[width=0.5\linewidth]{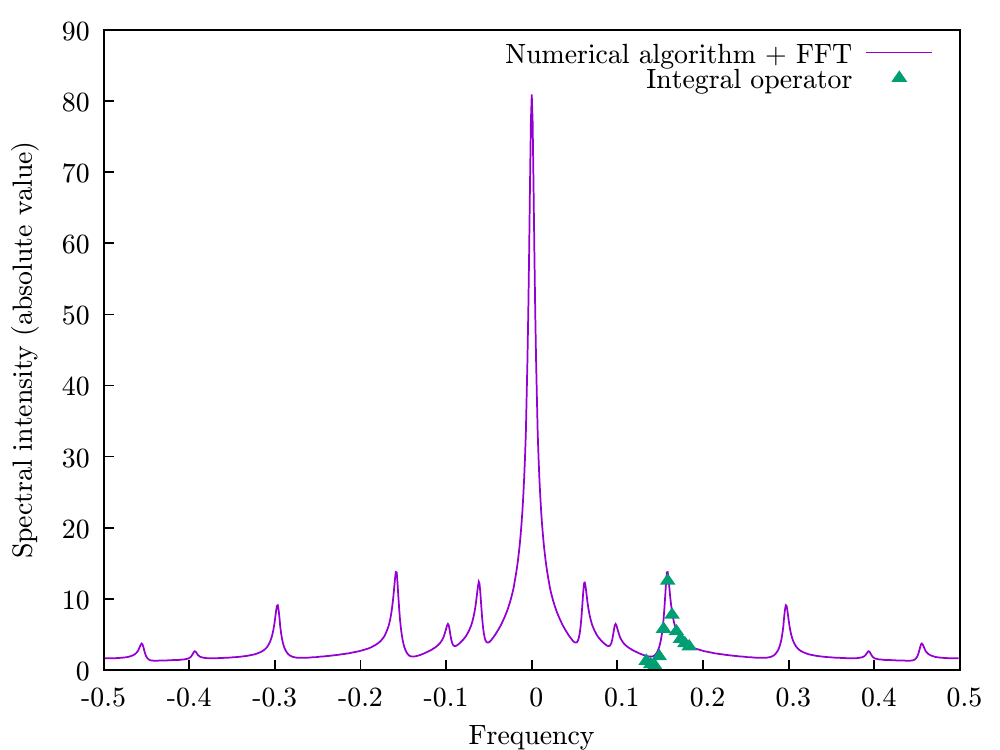}
	\caption{FFT of numerical time evolution of the expectation value of $S_N$ with a random Hamiltonian of size 4 compared with a few points computed with the integral operator.}
	\label{fig:test2}
\end{figure}

Figure \ref{fig:test2} shows instead the example of an application to Fourier transform of the time evolution generated by a random Hamiltonian of size 4. While it's not strictly necessary, in this example a damping time $\tau=50$ has also been used to broaden the peaks and represent the most general case. After performing the FFT of the numerically computed time evolution a peak is identified and a small window around it is explored with the integral operator. The results are in excellent agreement. Use for Fourier transform is not time efficient: in this example, the FFT on 1000 points has taken approximately $0.01\,s$, while each point computed with the integral operator has taken almost $45\,\mu s$. At this rate, if we had to compute the whole spectrum that way it would take almost 5 times longer. However, this method allows one to explore single values or small frequency windows offset from the origin, and in the case that is all that is needed it could result convenient.

\section{Conclusions}
A simple method that to our knowledge was not previously documented to compute exponentially weighted time integrals of an expectation value in quantum mechanics simulation has this proposed. This approach is shown to be significantly faster and more precise than traditional ones. Numerical simulations in the field of muon spectroscopy and nuclear magnetic resonance could potentially benefit from using this method in select cases of interest.

\section{Acknowledgments}

Thanks to Stewart Clark, Dominik Jochym, Barbara Montanari, Leonardo Bernasconi and Leandro Liborio for useful conversations. This work was supported by the Collaborative Computational Project for NMR Crystallography, funded by the EPSRC (UK) Grants EP/J010510/1 and EP/M022501/1.

\bibliographystyle{unsrt}
\bibliography{manuscript.bib}

\end{document}